\documentclass[12pt,superscriptaddress,floatfix]{iopart}
\expandafter\let\csname equation*\endcsname\relax
\expandafter\let\csname endequation*\endcsname\relax
\usepackage{color}
\usepackage{amsmath, amssymb, hyperref, multirow}
\usepackage[dvips]{graphicx}
\usepackage[]{subfigure}

\begin{document}

\title[Master equation approach to the subway passenger flow]
{Master equation approach to the intra-urban passenger flow and
application to the Metropolitan Seoul Subway system}

\author{Keumsook Lee$^1$\footnote{E-mail: kslee@sungshin.ac.kr}, Segun Goh$^2$, Jong Soo Park$^3$, Woo-Sung Jung$^4$, and M.Y. Choi$^2$\footnote{E-mail: mychoi@snu.ac.kr}}
\address{$^1$Department of Geography, Sungshin University,
Seoul 136-742, Korea}
\address{$^2$Department of Physics and Astronomy and Center for Theoretical Physics,
Seoul National University, Seoul 151-747, Korea}
\address{$^3$School of Information Technology, Sungshin University, Seoul 136-742, Korea}
\address{$^4$Department of Physics and Basic Science Research Institute,
Pohang University of Science and Technology, Pohang 790-784, Korea}

\begin{abstract}
The master equation approach is proposed to describe the evolution of passengers in a subway system.
With the transition rate constructed from simple geographical consideration, the evolution equation for the distribution of subway passengers is found to bear skew distributions including log-normal, Weibull, and power-law distributions.
This approach is then applied to the Metropolitan Seoul Subway system: Analysis of the trip data of all passengers in a day reveals that the data in most cases fit well to the log-normal distributions. Implications of the results are also discussed.
\end{abstract}

\pacs{89.75.Fb, 05.65.+b, 89.40.Bb}

\maketitle

\section{Introduction}

In a neural network, the functional network structure should depend on the anatomical structure of the network, which reflects the spatial distribution of neurons \cite{Hellwig00, Bullmore09}.
Similarly to such biological systems, physical space as well as mathematical topology is relevant also for spatial networks of social systems. This is reflected by the old proverb saying that out of sight, out of mind, in accord with which people tend to make their relationships in their neighborhood. Even in the case of the internet network as a distinctive social network, which is in itself completely independent of the geographic world, its physical realizations such as routers and links are located in the geographic real world \cite{Lakhina03}. Obviously, the spatial distance discussed here is not limited to the Euclidean distance and general geographic measure can be defined appropriately on a case-by-case basis. Particularly for the subway system, which provides a good example of the transportation network, geographic space affects the structure of the system in a critical way. For the analysis of traffic flows on such a transportation network, it is appropriate to introduce weighted networks \cite{Yook01, Barrat04}.

So far, various approaches to spatial networks have been developed \cite{Barthelemy10}.
In particular, emergence of skew distributions like power-law distributions, possibly with cut-offs,
or log-normal distributions was reported extensively in diverse systems including
maritime transportation \cite{Hu09}, London street networks \cite{Masucci09}, and passenger flows on subway networks
in Seoul \cite{subway} and in London \cite{Roth10}. Such skew distributions in spatial networks should naturally
emerge from the growth and evolution of elements of the systems. 
The Yule process or preferential attachment \cite{Barabasi99} is well known to generate a power-law distribution 
whereas the cut-off behavior in the power-law distribution can be attributed to 
dependence on the distance \cite{Barthelemy03} or geo-political constraints \cite{Guimera04}. 
Also the emergence of various kinds of networks or distributions may be controlled by the limitation rule 
on the addition of new links \cite{Amaral00}. 
These demonstrate that the structure of a system can be closely related to the time evolution of the system. 
Accordingly, examining the growth of a system, one may obtain crucial information on interesting characteristics 
of the system such as the size distribution.

In this paper, we study the time evolution of the passenger flow on a subway transportation network.
The number of passengers in a subway system evolves with the urban development. 
For example, if a new factory is  built or a financial company is relocated in the city,
the total number of passengers using the subway station near these facilities
or taking trips between the subway stations near these facilities and those near the residential area of laborers 
should grow.
The resulting time evolution of the number of passengers can be described conveniently by the master equation, which was proposed for growth \cite{master}.
Based on the master equation, where the transition rate is constructed from simple geographical consideration, we obtain the evolution equation for the distribution of subway passengers.
It is disclosed that the equation admits skew distributions including the log-normal and Weibull distributions as well as the power-law one.
We then consider the Metropolitan Seoul Subway (MSS) system, in which the trip trajectory data are available for all passengers in a day. Analysis of the huge data, consisting of more than $10^7$ trips, indeed reveals skew distributions. The origin of such distributions together with their implications is discussed.

This article consists of four sections: Section \ref{sec:master_eq} presents the master equation approach to the growth and branching processes. In Sec. \ref{sec:result} the MSS system is considered and the passenger data are analyzed, disclosing skew distributions.  Finally, Sec. \ref{sec:summary} discusses the obtained results and gives a summary.

\section{Master equation and skew distributions}
\label{sec:master_eq}

We consider the time evolution of the number of subway passengers $x$ using a given station or taking trips between two stations.
The passenger number $x$ will change to $x^{\prime}$ when a factory or an apartment
complex is built; this will be called an event.
The amount of the change, $\Delta x \equiv x^{\prime}-x$ is expected to depend on the present number $x$.  Assuming the rich-get-richer principle, we take $\Delta x$ to be proportional to $x$: $\Delta x =bx$ with the growth factor $b$.
The most simple form of the transition rate then reads
\begin{equation}
\label{eq:tran_rate}
\omega (x \to x^{\prime})=\lambda \delta (x^{\prime}-x -bx)
\end{equation}
where $\lambda$ is the growth rate or the occurrence rate of the events affecting $x$.

With the transition rate above, we now establish the master equation for the probability of the passenger numbers.
Suppose that there are $N$ elements, which correspond to either $N$ subway stations or $N$ links between any pairs of subway stations. (Note that links are associated with passengers taking trips between origin and destination stations, and there is a link not only between two adjacent stations on one subway line but also between any two stations in the whole subway system.)
The number of passengers of the $i$th element is denoted by $x_i$.
We then follow Ref. \cite{master} to write the master equation for the probability
$P(x_1 , x_2 ,\cdots,x_N ;t)$
for the subway system in specific configuration $\{x_1,x_2,\cdots,x_N\}$ at time $t$: 
\begin{align}
\label{eq:mastereq}
\frac{\mathrm{d}}{\mathrm{d} t} P(x_1,\cdots,& x_N;t)
=\sum_{j} \int \mathrm{d} x_j^{\prime} \,
[\omega (x_j^{\prime}\to x_j ) P(x_1,\cdots,x_j^{\prime},\cdots,x_N;t) \nonumber \\
&-\omega (x_j \to x_j^{\prime}) P(x_1,\cdots,x_j,\cdots,x_N;t) ]
\end{align}
where the transition rate $\omega$ is given by Eq. (\ref{eq:tran_rate}).

Here we introduce the distribution function $f(x,t)$ of the passengers \cite{master},
which is related to the system configuration probability via
\begin{equation}
f(x,t)=\frac{1}{N}\int \mathrm{d} x_1 \cdots \mathrm{d} x_N \sum_{i} \delta (x_i -x) P(x_1,\cdots,x_N;t).
\end{equation}
Equation~(\ref{eq:mastereq}) then takes the form
\begin{align}
\frac{\partial}{\partial t} f(x,t)
 = &\frac{\lambda}{N}\sum_i \int \mathrm{d}x_1 \cdots \mathrm{d}x_N \mathrm{d}x_i^{\prime} \,
   \delta (x_i -x)
	[\delta ((1{+}b)x_i^{\prime}-x_i )P(x_1,\cdots,x_i^{\prime},\cdots,x_N;t) \nonumber \\
  &-\delta (x_i^{\prime}-(1{+}b)x_i)P(x_1,\cdots,x_i,\cdots,x_N;t)],
\end{align}
where the summation over the index $j$ has been performed.
Integrating over $x_i^{\prime}$ with care to the accurate normalization,
we arrive at the equation
\begin{equation}
\label{eq:evolutioneq}
\frac{\partial}{\partial t}f(x,t)=-\lambda f(x,t)+\frac{\lambda}{1+b}\,f\left(\frac{x}{1{+}b}\right),
\end{equation}
which describes the time evolution of the distribution function.

It is straightforward to compute the stationary solution of Eq. (\ref{eq:evolutioneq}).
Setting $f(x,t) = f_s (x)$ and the left-hand side of Eq. (\ref{eq:evolutioneq}) equal to zero,
we obtain
\begin{equation}
f_s (x) \sim x^{-1} .
\end{equation}
Namely, Eq. (\ref{eq:evolutioneq}) admits the trivial stationary solution, which is the power-law distribution with the exponent unity.

To probe the evolving (time-dependent) solution of Eq. (\ref{eq:evolutioneq}), we consider the passenger number $x$ on the logarithmic scale.
On the linear scale, the number is determined by multiplying appropriate times by a random variable $X$ which takes the value $1$ or $(1 {+} b)$ independently (and obviously has a finite second moment).
Then on the logarithmic scale, it is given by adding independently $0$ or $\ln (1{+}b)$ appropriate times,
which, according to the central limit theorem, results in a Gaussian distribution (for logarithms of $x$).
This can be obtained directly from the time evolution equation by changing the variable
from $x$ to $X \equiv \ln x$. In terms of the corresponding distribution function $F(X,t) = e^{X} f(e^X , t)$, Eq. (\ref{eq:evolutioneq}) simply reads
\begin{equation}
\frac{\partial}{\partial t}F(X,t) =-\lambda F(X,t) +\lambda F(X-a,t)
\end{equation}
with $a \equiv \ln{(1+b)}$, which describes exactly the random walk and yields the Gaussian (normal)
distribution as the asymptotic solution.
This corresponds to a log-normal distribution for $x$ as the solution of the original model described by Eq. (\ref{eq:evolutioneq}).
Indeed direct substitution confirms that Eq. (\ref{eq:evolutioneq}) bears the log-normal distribution asymptotically, with time-dependent parameters.
Specifically, inserting the log-normal distribution
\begin{equation}
\label{eq:lognormal}
f(x,t)=\frac{1}{\sqrt{2\pi}\sigma x}\mathrm{e}^{-\frac{1}{2\sigma^2}(\ln{x}-\mu)^2}
\end{equation}
in Eq. (\ref{eq:evolutioneq}), we have
\begin{equation}
\label{eq:log_evol_eq}
\frac{1}{\lambda}\left[ -\frac{\dot{\sigma}}{\sigma} + \frac{\dot{\mu}}{\sigma^2}(\ln{x}-\mu )
 +\frac{\dot{\sigma}}{\sigma^3}(\ln{x}-\mu )^2 \right]
 = \exp\left\{\frac{\ln{(1{+}b)}}{2\sigma^2}[2\ln{x}-\ln{(1{+}b)}-2\mu ] \right\} -1 .
\end{equation}
Expanding the right-hand side with respect to $\sigma^{-1}$
gives the time evolution of the mean and deviation parameters $\mu$ and $\sigma$ of the log-normal distribution, obtained to the order of $\sigma^{-2}$:
\begin{eqnarray}
\label{eq:sigma}
\mu &=& a \lambda t \nonumber \\
\sigma &=& a \sqrt{\lambda t}
\end{eqnarray}
with $a\equiv \ln (1{+}b)$.
Note that $\lambda t$, product of the growth rate $\lambda$ and elapsed time $t$,
corresponds to the growth frequency, i.e., the number of events occurred.
It is also straightforward to generalize Eq. (\ref{eq:evolutioneq}) for
the system where there occur various types of event described by different values of $\lambda$ and $b$.
For example, in the case of two types of event characterized by $(\lambda_1, b_1)$ and $(\lambda_2, b_2)$, the evolution equation reads
\begin{equation}
\label{eq:evolutioneq_branch}
\frac{\partial}{\partial t}f(x,t)=
-\lambda_1 f(x,t)+\frac{\lambda_1}{1+b_1}f\left(\frac{x}{1{+}b_1}\right)
-\lambda_2 f(x,t)+\frac{\lambda_2}{1+b_2}f\left(\frac{x}{1{+}b_2}\right).
\end{equation}
The log-normal distribution in Eq. \ref{eq:lognormal} again makes a solution,
where the mean and the deviation are given by
$\mu = (a_1 \lambda_1 +a_2 \lambda_2 )t$ and
$\sigma = \sqrt{(a_1^2 \lambda_1 + a_2^2 \lambda_2 ) t}$
with
$a_1 \equiv \ln (1{+}b_1 )$ and $a_2 \equiv \ln (1{+}b_2 )$.
Fitting the data at given time to Eq. (\ref{eq:lognormal}),
we can surmise the time evolution of the number of passengers
and predict the distribution in the future.

Heretofore the subway system has been considered as fixed.
In reality, the number of stations or links in a subway system can change with time
as construction of the system proceeds.
Usually, a new subway station is added as a member of a newly constructed subway line,
which connects, like most existing lines, downtowns and suburbs or business districts and residential areas.
In consequence, the distribution of passengers due to newly constructed stations, i.e., passengers using new stations and links, is expected to be similar to that of existing stations and links.
In case that the interactions between existing stations and new ones are negligible,
Eq. (\ref{eq:evolutioneq}) is also applicable even when new stations and links are included.
It is, however, conceivable that increases in the numbers of passengers using new stations or links are related to the numbers of passengers of existing stations.
Namely, additional stations and links in the subway system should induce redistribution of passengers as well as influx of new passengers (e.g., from newly accessible parts of the city).
Such redistribution corresponds to the branching process, which can be described similarly to the growth process.
In particular, the simple branching process including just one scale of branching like binary fission is described by the evolution equation in Eq. (\ref{eq:evolutioneq_branch}) with $b_2$ less than zero. (Note that a negative value of $b$ implies a decrease of passengers.)
Whereas the terms involving $\lambda_1$ and $b_1$ represent the growth as the terms in Eq.~(\ref{eq:evolutioneq}),
those involving $\lambda_2$, related to the construction rate of new stations,
and $b_2$, reduction factor, correspond to the branching process.

Equation~(\ref{eq:evolutioneq_branch}) admits the Weibull distribution of the form
\begin{equation}
\label{eq:Wei_dist}
f(x,t)=\frac{\gamma}{x} \left( \frac{x}{\eta} \right)^{\gamma} \mathrm{e}^{-(x/\eta)^{\gamma}}
\end{equation}
in addition to the log-normal distribution.
Inserting Eq. (\ref{eq:Wei_dist}) and expanding as before Eq.~(\ref{eq:log_evol_eq})
indeed show that the Weibull distribution is an asymptotic solution,
with the shape and scale parameters given by
\begin{eqnarray}
\label{eq:gamma}
\gamma &= \frac{1}{\sqrt{(a_1^2 \lambda_1 +a_2^2 \lambda_2)t}}, \nonumber \\
\eta &= \exp{[(a_1 \lambda_1 +a_2 \lambda_2)t]} .
\end{eqnarray}
While the log-normal distribution has bilateral symmetry after transformed to the normal distribution,
the Weibull distribution still has asymmetry, skewness leaned to the front and having a light tail,
as the consequence of the branching process.
%

%
To conclude, emergence of the log-normal or Weibull distribution depends on the manner of production of new elements, i.e., construction of new subway lines. In case that only existing elements grow independently by given growth factor with no elements created newly or that the weight and the strength distributions of new elements are similar to those of existing stations and links (corresponding to the self-size production), the size distribution of the system is expected to follow the log-normal distribution. On the other hand, if new elements are created from fission and the size of newly created elements is determined by the redistribution of the size of pre-existing elements (corresponding to the branching process), the Weibull distribution is expected to emerge. Detailed results for these production rules can be found in Ref.~\cite{skew}.

Further, these processes can affect alternatively the growth of the system in a complementary manner.
For example, one may consider the case that a portion of the passengers at a subway station consists of entirely new users of the subway system, i.e., incomers to the city or those who have used other means of transportation, and the other portion consists of passengers from other subway stations. The resultant distribution of weights or strengths will then lie somewhere in between the log-normal distribution
and the Weibull one.

\section{Metropolitan Seoul Subway system}
\label{sec:result}

In this section, equipped with the results of Sec.~\ref{sec:master_eq}, we analyze the passenger data of the MSS system.
We first define such terms as the weight, directed weight, and strength, following Ref.~\cite{subway}: The weight $w$ of a link between a pair of subway stations is defined to be the total number of passengers taking trips between the two stations, flowing in both directions, whereas
the directed weight is the number of passengers in one direction, i.e., from one station (origin) to the other (destination).
The (total) strength $s$ of a subway station is defined to be the total number of passengers using the station,
so that the strength is equal to the sum of the weights of the links between the station and others.
We also define the strength of departure and that of arrival as the numbers of passengers departing from and arriving at the station.
Obviously, the sum of the departure and arrival strengths gives the total strength.

There are 357 subway stations and 127,092 links between stations in the MSS system data used in this study,
giving the data set to be handled: $X=\{x_1 ,\cdots, x_n \}$ with $n=357$ for strengths and $n=127,092$ for weights. We deal with one-day data, which are tremendous in size, consisting of more then $10^7$ passenger trips (including bus trips). For convenience, we divide a day into three time zones: morning (4 am to 10 am), afternoon (11 am to 4 pm), and evening (5 pm to 1 am), and separate the data accordingly, e.g., departure in the morning, arrival in the afternoon, etc.
More specifically, 4,907,541 passengers depart from and arrive at stations in the whole day: 1,671,891 passengers depart in the morning, 1,128,482 passengers in the afternoon, and 2,107,168 passengers in the evening. Similarly, 1,585,358 passengers arrive at stations in the morning, 1,082,196 passengers in the afternoon, and 2,239,987 passengers in the evening. Those arrivals and departures are counted independently,
which results in that arrival and departure numbers in the strength data do not match with each other.
This is due to the passengers who extend their trips beyond a single time zone, for example,
getting on the subway in the morning and getting off in the afternoon.
On the other hand, weights involve the whole duration of each trip, from the origin to the destination,
and those passengers whose trips are extended beyond one time zone are excluded in the weight data.

\begin{figure}
\centering
\subfigure[\ Morning]{
	\includegraphics[width=0.7\textwidth, angle=0]{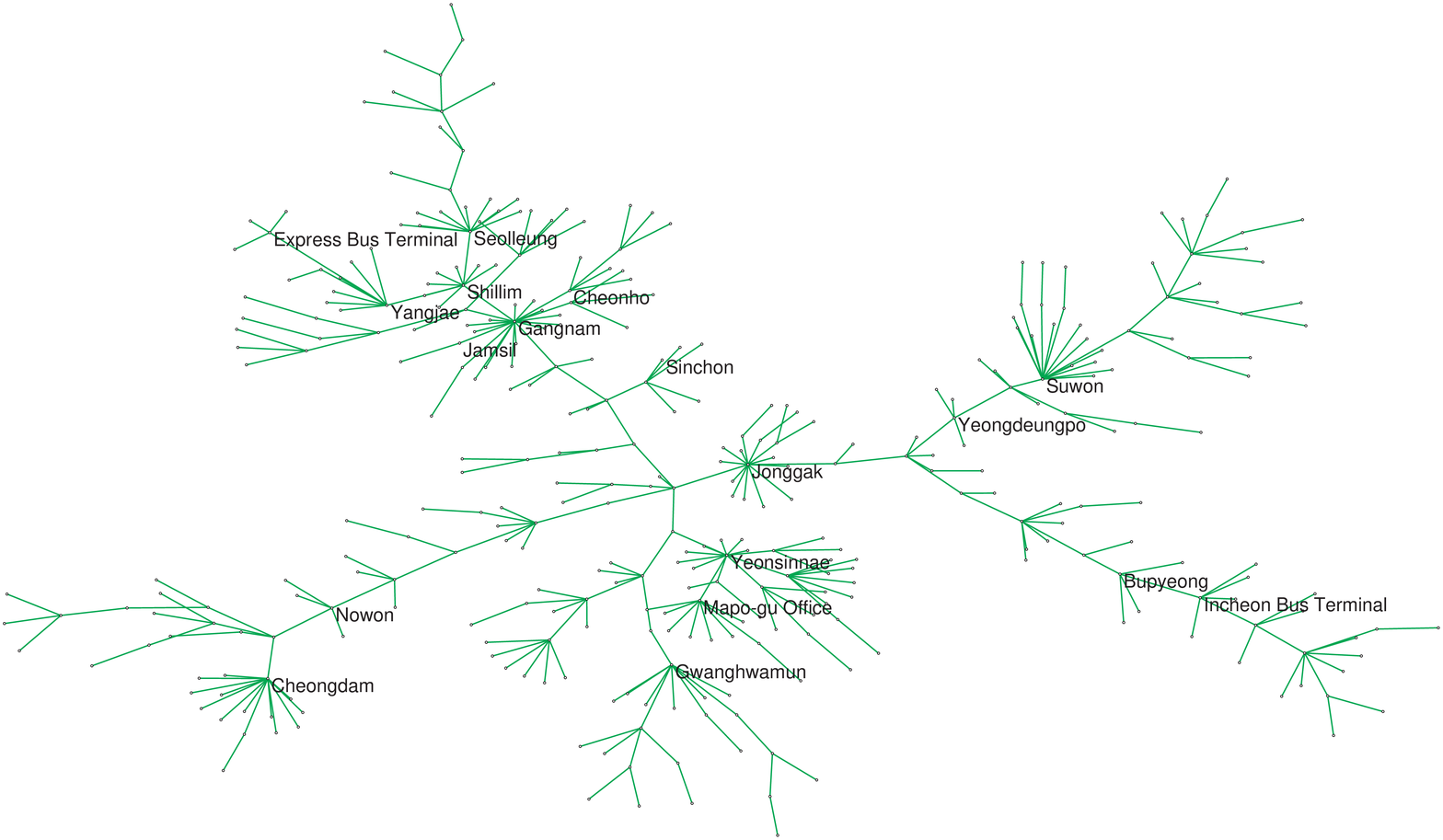}}
\subfigure[\ Afternoon]{
	\includegraphics[width=0.7\textwidth, angle=0]{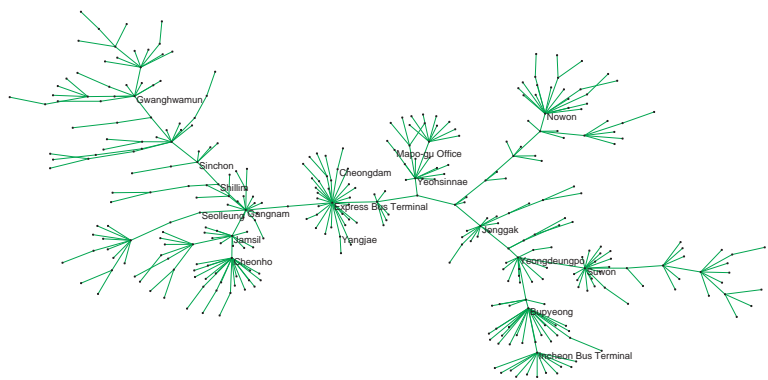}}
\subfigure[\ Evening]{
	\includegraphics[width=0.7\textwidth, angle=0]{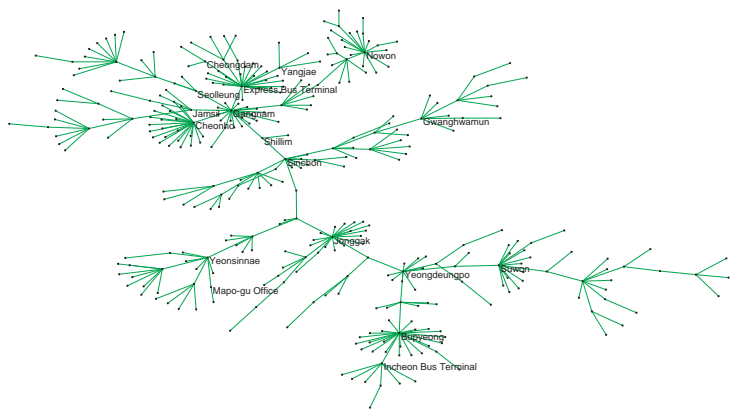}}
\caption{
Maximum spanning trees of passenger flows in the Metropolitan Seoul Subway System.
Names of some hub stations are provided for reference.
}
\label{f:mst}
\end{figure}

We first consider the spanning trees of passenger flows, where nodes represent subway stations,
in the morning, afternoon and evening.  The minimum spanning tree, which is widely employed to investigate the structure of a complex network,
is constructed with weights not larger than those of other possible spanning tree.
However, in the subway system, the links which have larger weights (passenger flows) are more important;
those links make up the so-called {\it maximum} spanning tree (MST) \cite{subway}.
Figure \ref{f:mst} presents the MSTs of passenger flows, built from the weight data in the morning,
afternoon, and evening.
The different MSTs in each time zone manifest that the hub structure changes with time in a day,
which was not addressed in the previous study \cite{subway}.
In the morning, large flows due to commuting passengers should occur
between a number of (suburban) residential areas and a few (downtown) business districts;
thus the latter appear as hubs, characterized by large degrees (numbers of connected links)
in the MST.
If the commuting pattern in the evening were the opposite to that in the morning,
namely, if most passengers returned to their origins (residential areas) directly in the evening,
the resulting MST would be the same as that in the morning,
since the MST, constructed with (undirected) weights,
does not discern the flow direction on each link.
In reality, however, passengers tend to diffuse in the evening, rather than to return to
their origins. In consequence the MST displays different structure,
with more hubs usually of less degrees.

\begin{figure}
\centering
\includegraphics[width=0.5\textwidth, angle=0]{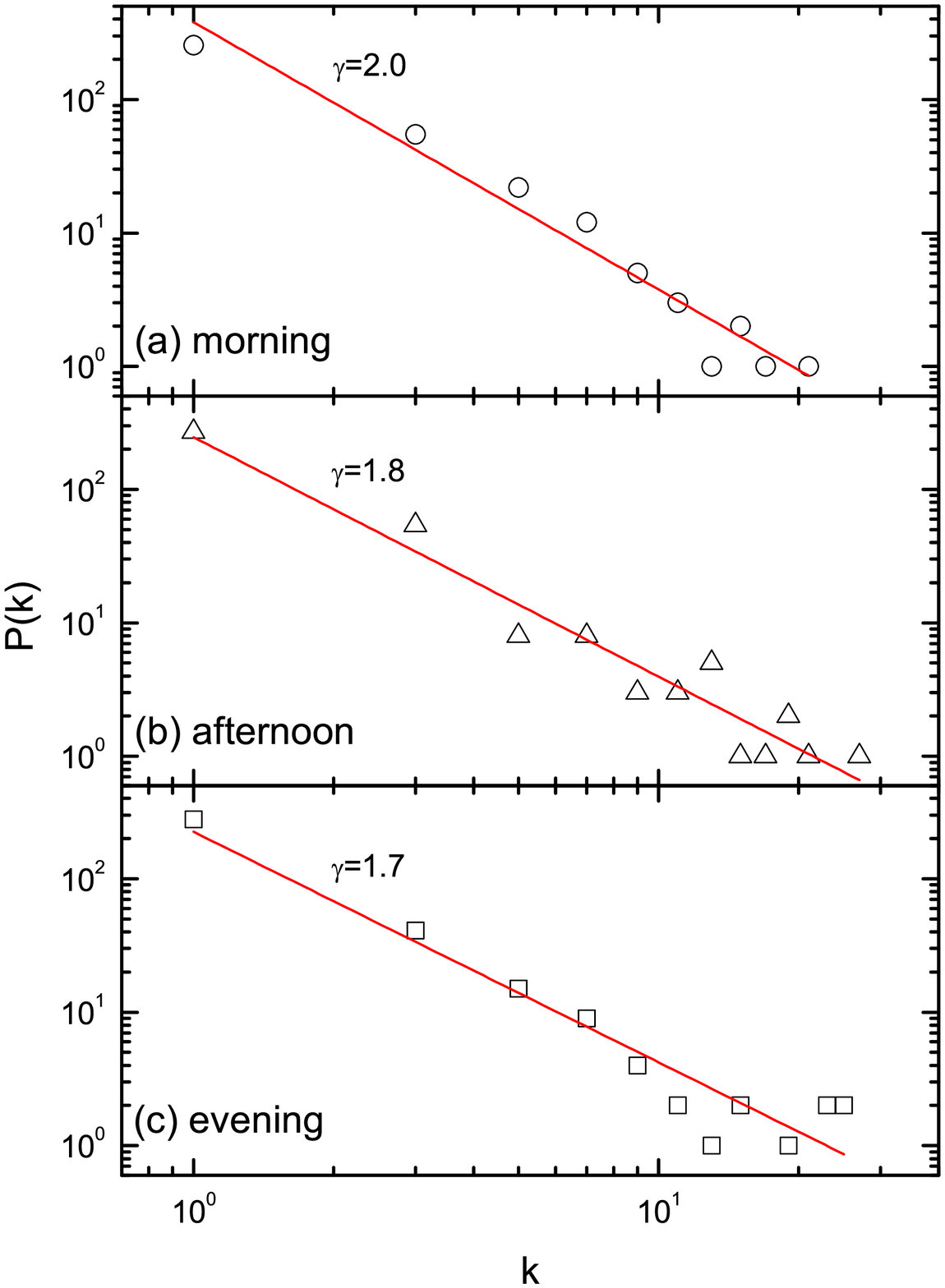}
\caption{
Probability distribution of the degree $k$ of the maximum spanning tree in Fig. \ref{f:mst}, plotted on the log-log scale. Lines serve as guides to the eye.
}
\label{f:degree}
\end{figure}

Counting the degree $k$ of each node (station) in the MST, we obtain the degree distribution $P(k)$,
which is shown in Fig. \ref{f:degree}.
We use the least-squares fitting, and observe the power-law behavior: $P(k)\sim k^{-\gamma}$ with exponent $\gamma \approx 2.0$, $1.8$, and $1.7$ for morning, afternoon, and evening flows, respectively. Note the slight variations of the exponent depending on the time zone.

As suggested in Sec.~\ref{sec:master_eq},
passenger data in the MSS system are fitted to the log-normal distribution and the Weibull distribution.
Both the least-squares fitting and the likelihood test method are used to
obtain values of $(\mu, \sigma)$ for log-normal distributions
and $(\gamma, \eta)$ for Weibull distributions;
then reduced $\chi^2$ and log likelihood function $\ln{L}$ are computed
to recognize which distribution gives better fitting of the data of MSS system.
Note that $\chi^2$, given by the sum of the squared differences between the data points and the fitted distribution, is widely used to test justification of the fit.
The likelihood function $L$ is defined to be~\cite{Hogg95}
\begin{equation}
L(\alpha, \beta) \equiv \prod_{i=1}^{n}f(x_i ; \alpha,\beta)
\end{equation}
where $\alpha$ and $\beta$ are parameters of the distribution function $f(x_i ; \alpha,\beta)$ to test,
and $x_i$'s are elements of the data set $X$.
We first find the maximum likelihood estimators $\hat{\alpha}$ and $\hat{\beta}$, which lead to the maximum value of the likelihood function.
Among the tested distributions,
the very distribution with the largest maximized likelihood value calculated from the maximum likelihood estimators
is taken to be the best distribution describing the data.
Validity of the likelihood test may be naively understood
if the conservation normalization is proposed as the constraint~\cite{Hogg95}.

Here we remark that there exist links with zero weight although all stations have finite strengths.
In general, links with zero weight are categorized into two groups:
One includes links connecting stations which are very far from each other.
In this case, passengers would rather take trips to the sub-center of the city closer to the origin.
The other consists of links involving several transfers.
Namely, transfer to a different subway line causes inconvenience and tends to be avoided; it is likely to use other transportation modes, e.g., cars and buses.
These features do not change with the growth of the city or the evolution of the subway system.
In addition, the number of passengers of an element, once it is zero, may not grow in our model,
which allows one to exclude such links with zero weight in analyzing the data.

\begin{figure}
\centering
	\includegraphics[width=0.33\textwidth, angle=270]{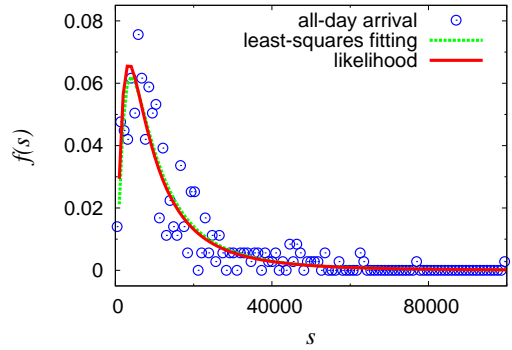}
\caption{
All-day arrival strength data, plotted in circles, together with the fitted log-normal distributions obtained by means of the likelihood method (red solid line) and least-squares fitting (green dotted line).
Both methods give similar values of the parameters $\mu$ and $\sigma$ within the error bar.
}
\label{f:strength_chi_like}
\end{figure}

\begin{figure}
\centering
\subfigure[\ Morning, arrival]{
	\includegraphics[width=0.33\textwidth, angle=270]{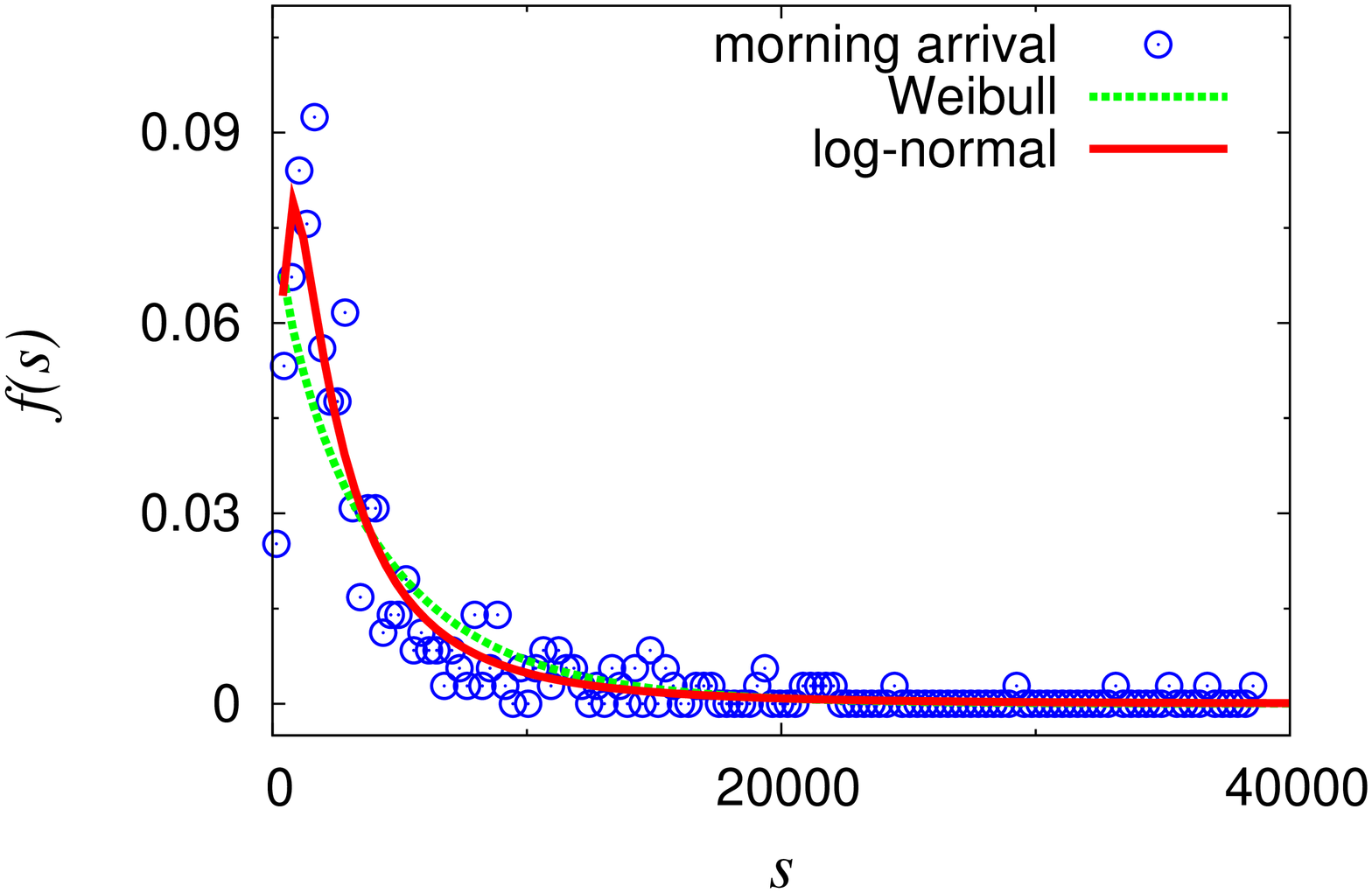}}
\subfigure[\ Evening, departure]{
	\includegraphics[width=0.33\textwidth, angle=270]{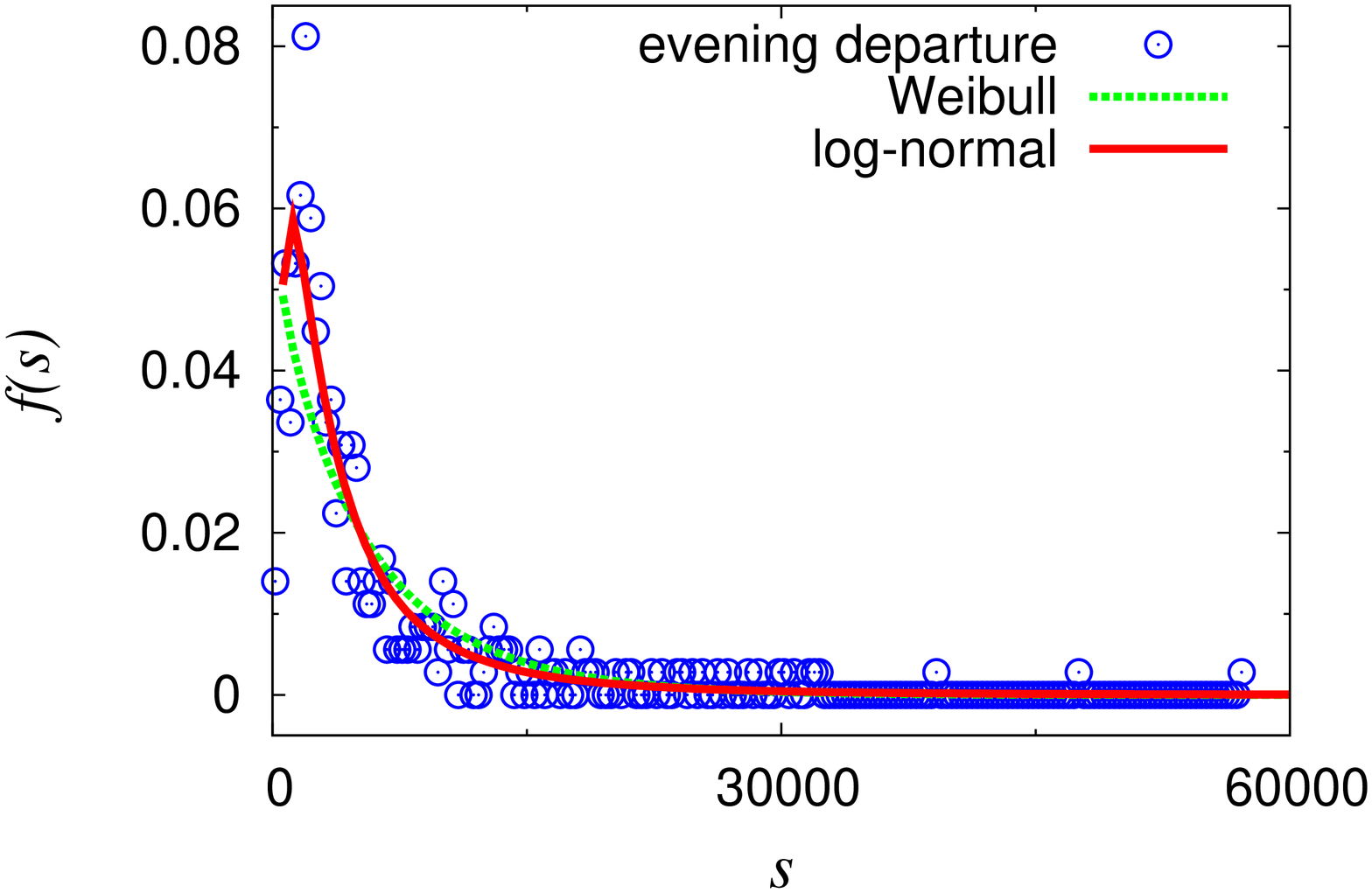}}
\subfigure[\ Afternoon, total]{
	\includegraphics[width=0.33\textwidth, angle=270]{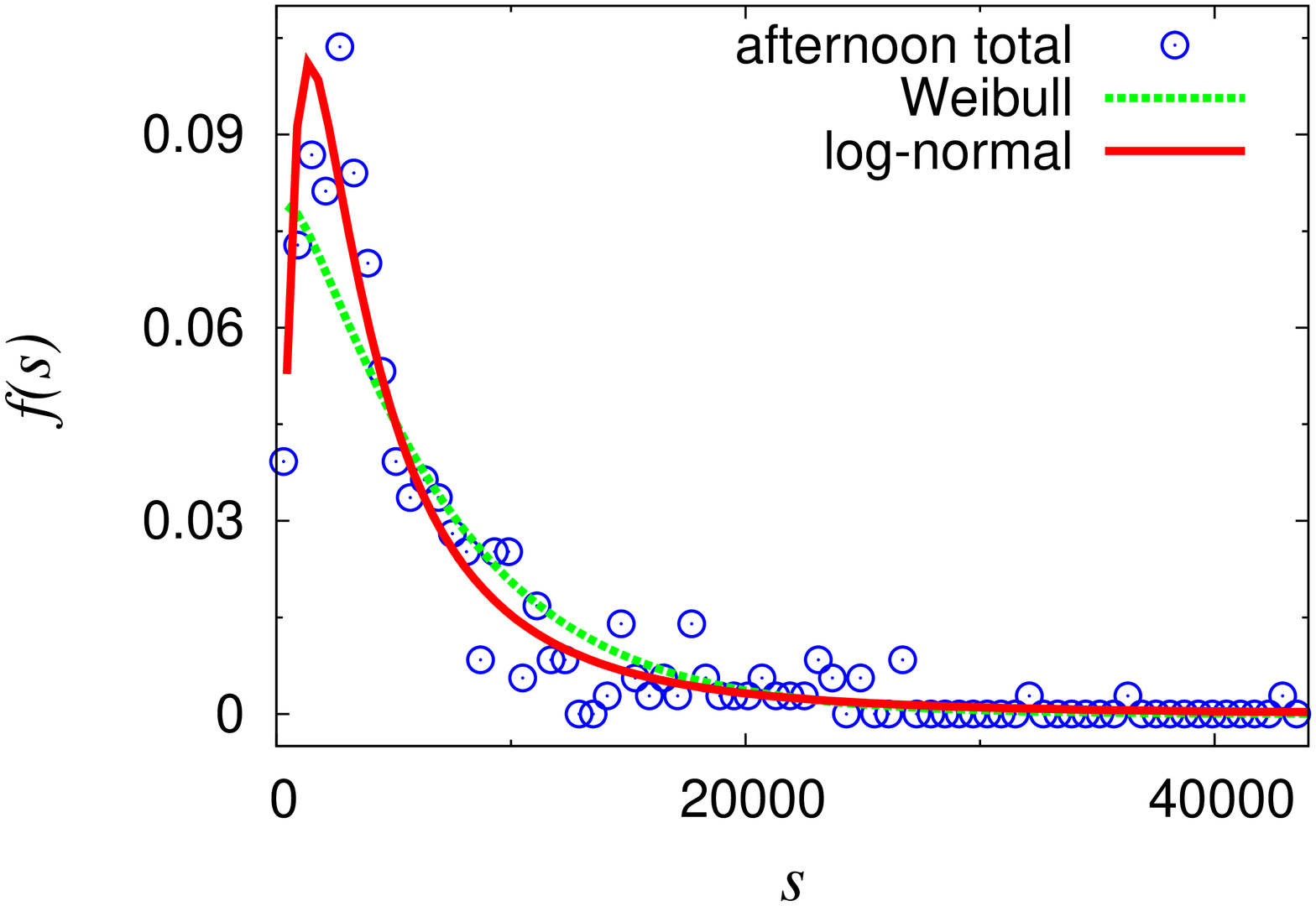}}
\subfigure[\ All day, total]{
	\includegraphics[width=0.33\textwidth, angle=270]{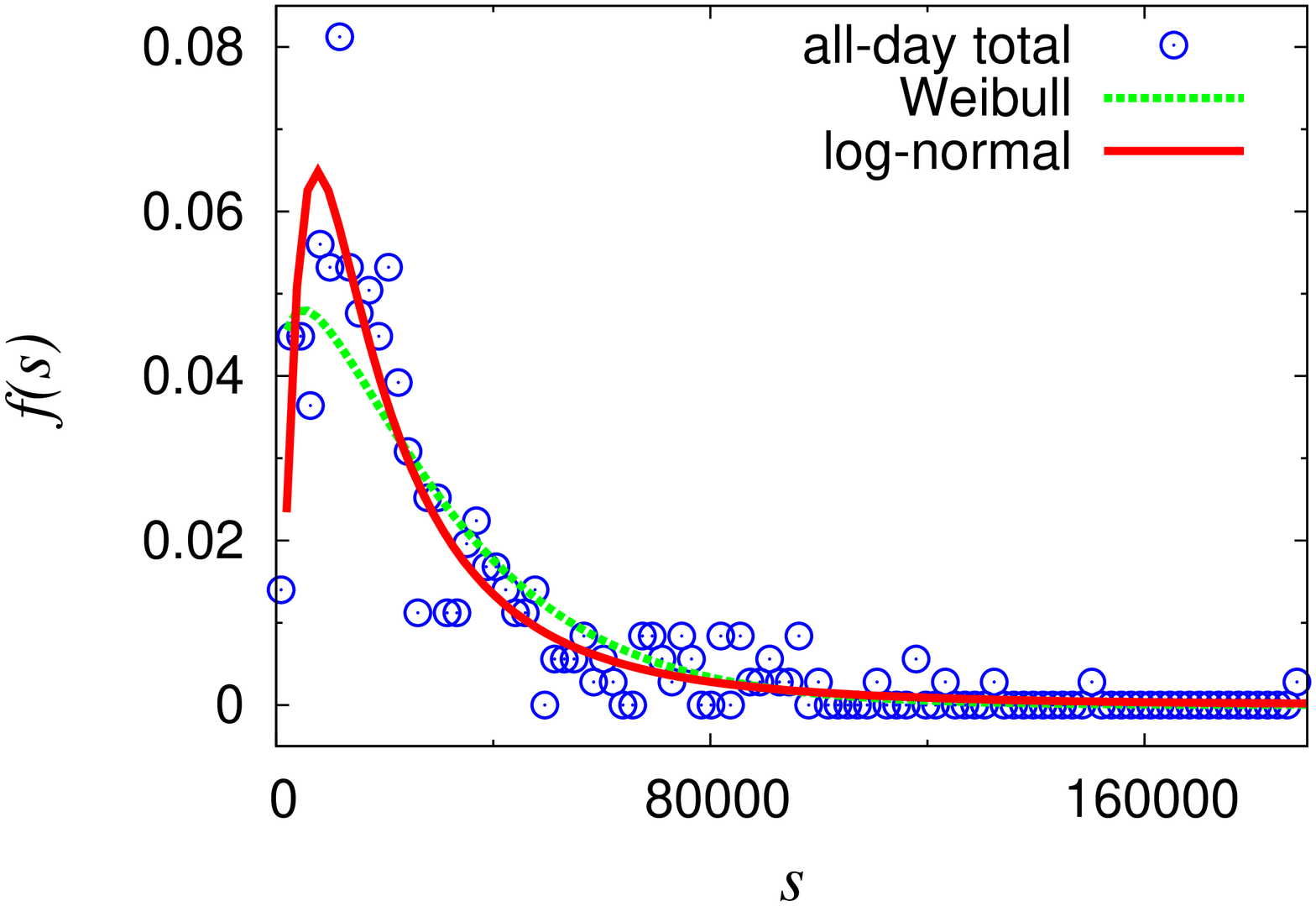}}
\caption{
Strength data in various cases, plotted in circles, are shown to fit well to log-normal distributions.
Both the log-normal distributions (red solid lines) and Weibull distributions (green dotted lines) have been obtained by mens of the likelihood method.
}
\label{f:strength_lognormal}
\end{figure}

\begin{figure}
\centering
\subfigure[\ Morning, departure]{
	\includegraphics[width=0.33\textwidth, angle=270]{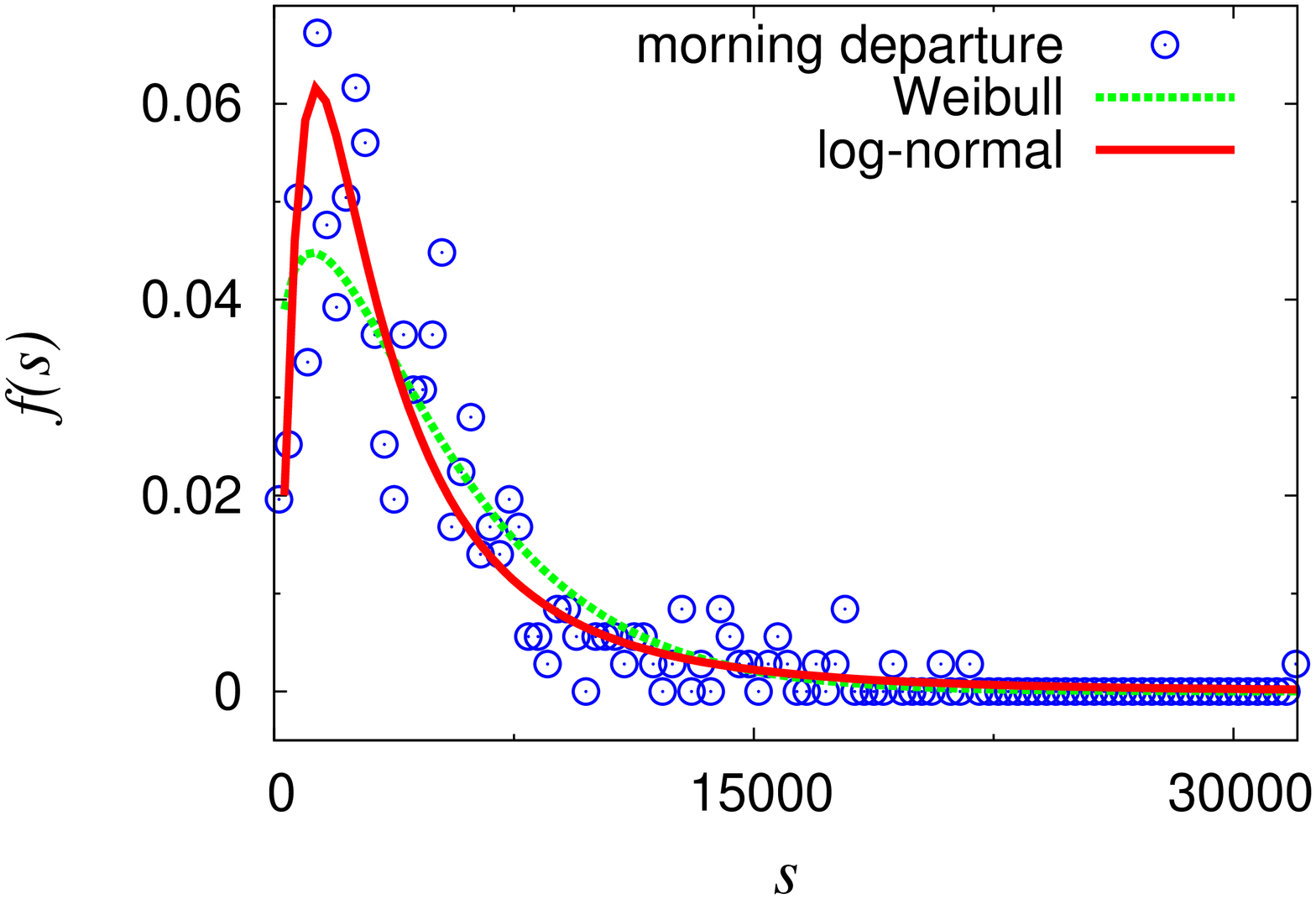}}
\subfigure[\ Evening, arrival]{
	\includegraphics[width=0.33\textwidth, angle=270]{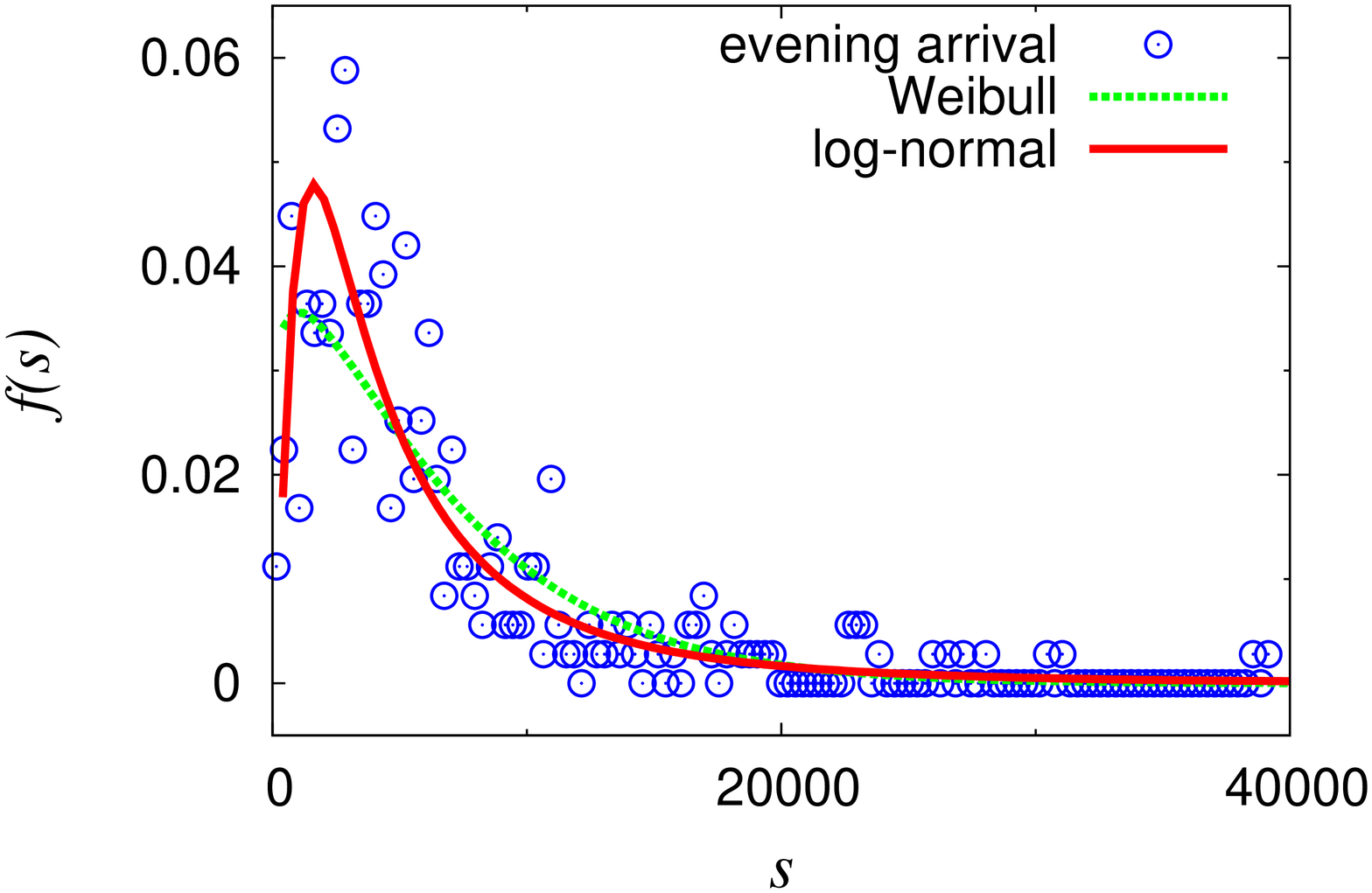}}
\caption{
Strength data fitted somewhat better to Weibull distributions.
Both log-normal and Weibull distributions have been obtained by means of the likelihood method.
Symbols are the same as those in Fig. \ref{f:strength_lognormal}.
}
\label{f:strength_Weibull}
\end{figure}

\begin{figure}
\centering
\includegraphics[width=0.33\textwidth, angle=270]{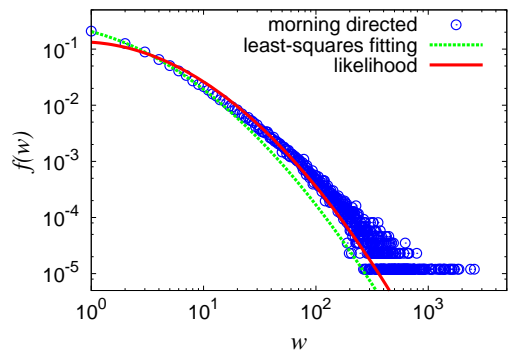}
\caption{
Data of the directed weights in the morning on the logarithmic scale, together with log-normal distributions obtained by means of the likelihood method and least-squares fitting.
They give slightly different values of the parameters.
In general, the likelihood test gives better results except for the region of small weights.
Symbols are the same as those in Fig. \ref{f:strength_chi_like}.
}
\label{f:weight_chi_like}
\end{figure}

\begin{figure}
\subfigure[\ All day non-directed]{
	\includegraphics[width=0.33\textwidth, angle=270]{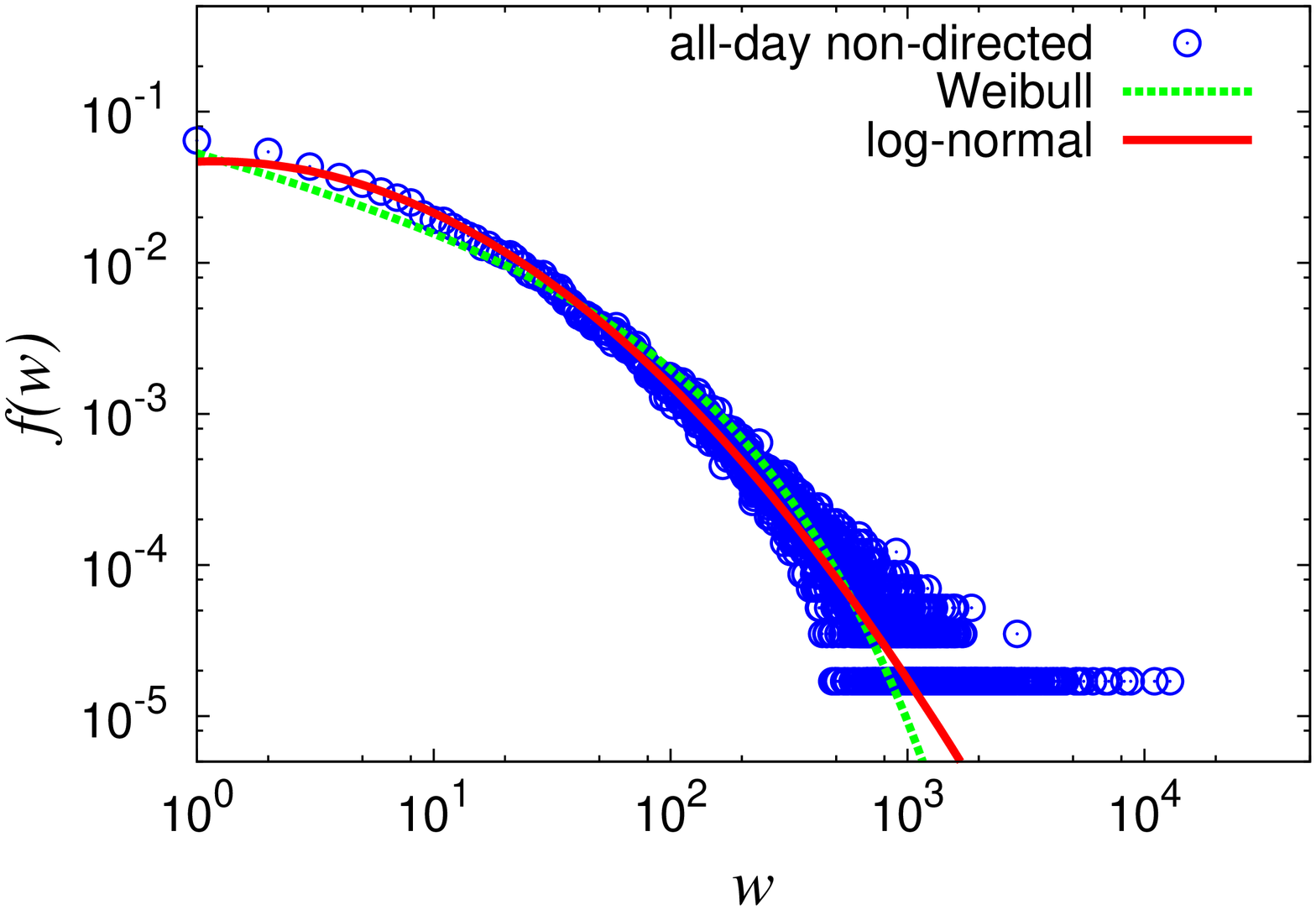}}
\subfigure[\ All day directed]{
	\includegraphics[width=0.33\textwidth, angle=270]{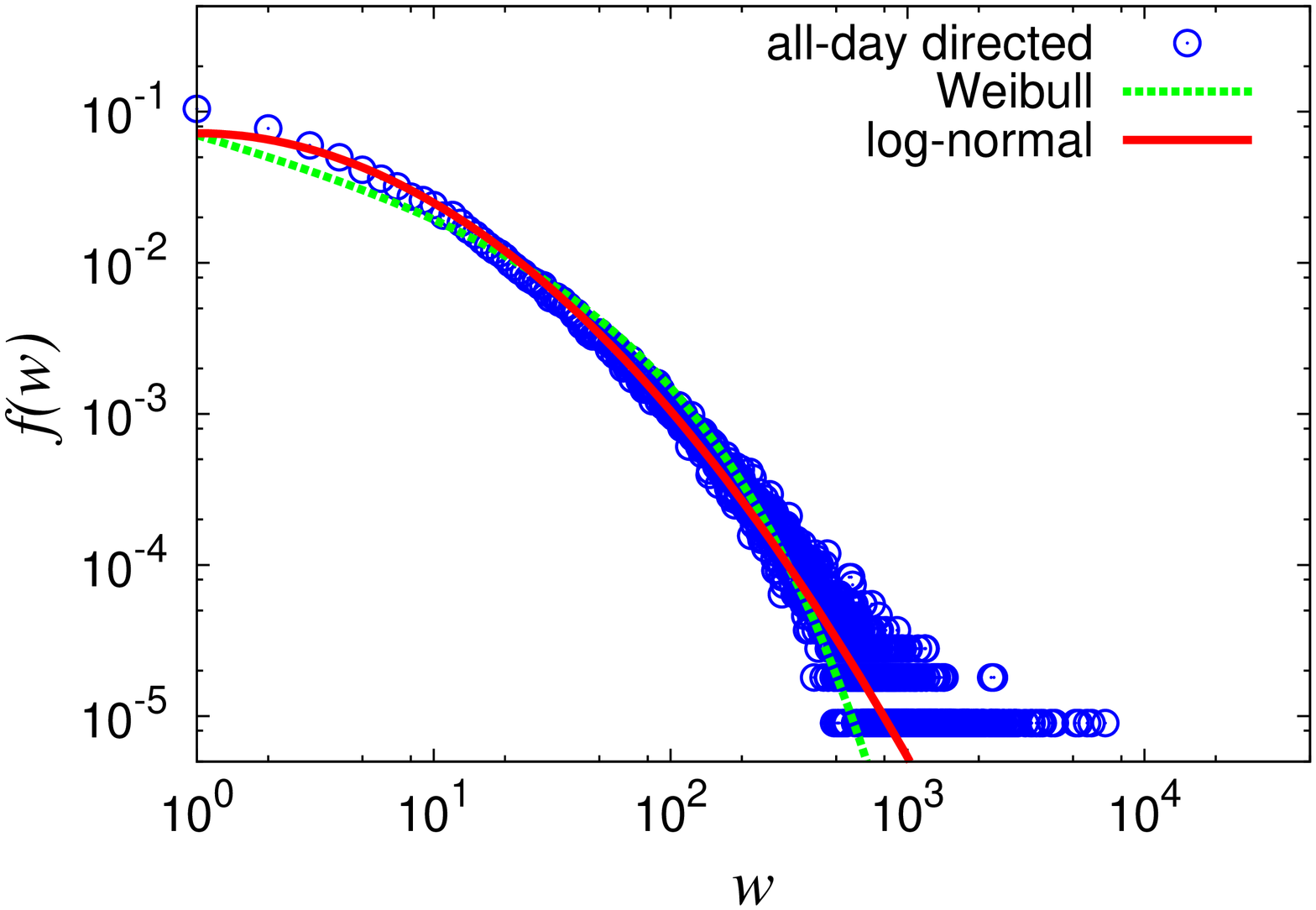}}
\subfigure[\ Afternoon non-directed]{
	\includegraphics[width=0.33\textwidth, angle=270]{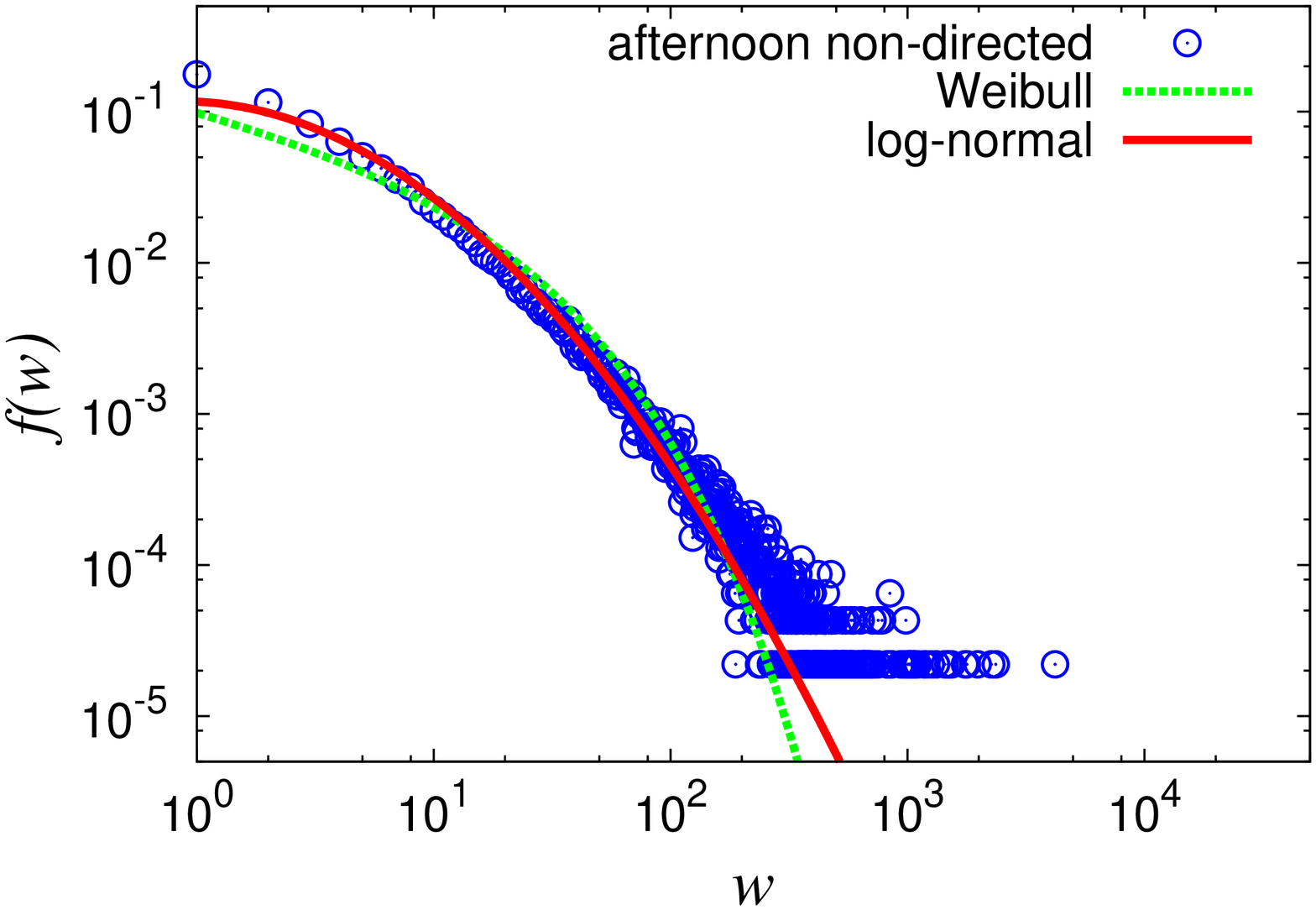}}
\subfigure[\ Evening directed]{
	\includegraphics[width=0.33\textwidth, angle=270]{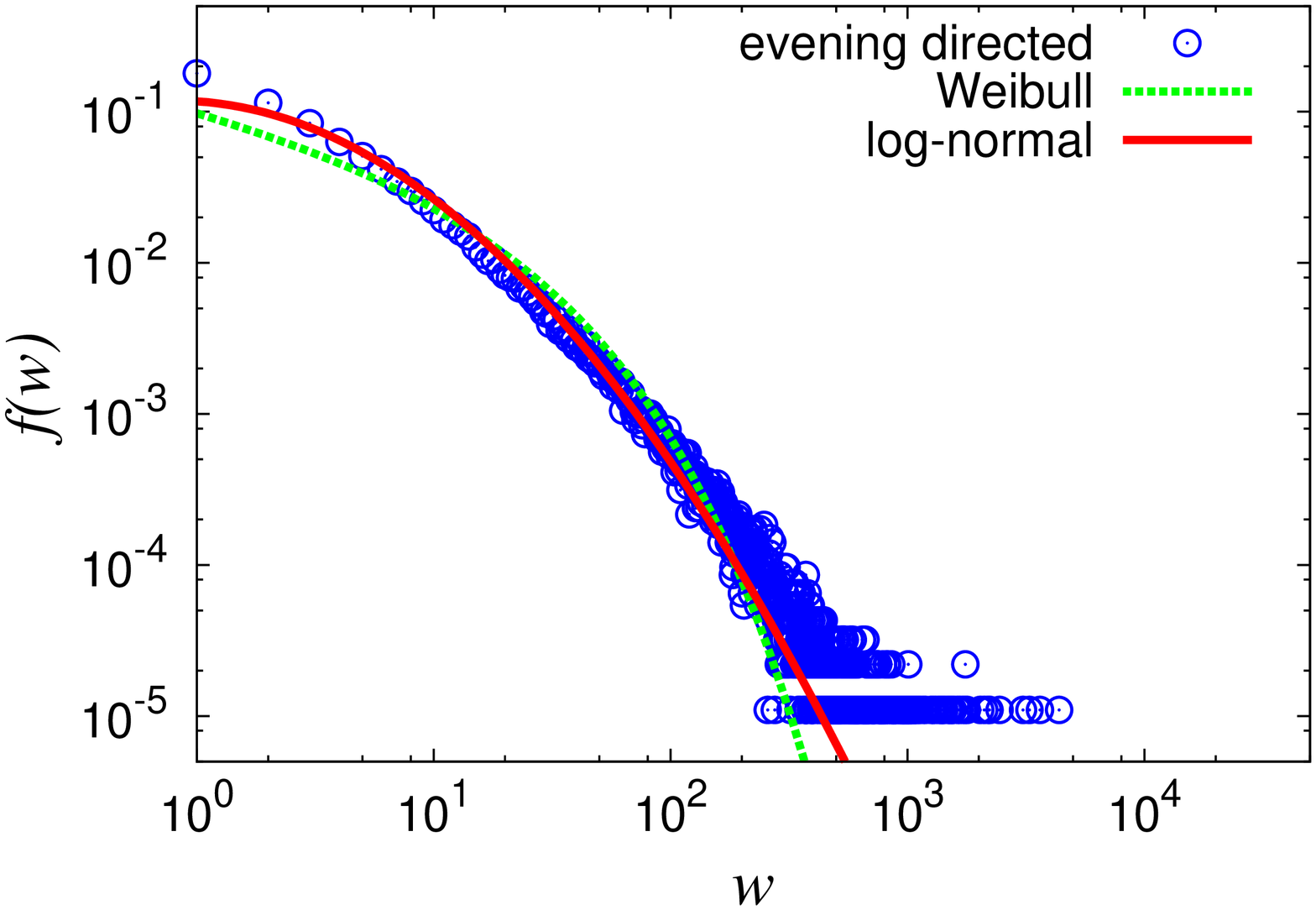}}
\caption{
Various weight data on the logarithmic scale.
All data fit well to log-normal distributions.
Symbols are the same as those in Fig. \ref{f:strength_lognormal}.
}
\label{f:weight}
\end{figure}

Computed values of chi square $\chi^2$ and log of the likelihood function $\ln L$
are given in Table~\ref{t:chi_lnL}.
In most cases, the least-squares fitting and the likelihood test agree with each other on the fitted distribution.
In case that they disagree, the difference in the likelihood value as well as in the $\chi^2$ value of log-normal and Weibull distributions
is small compared with the cases displaying agreed results.
This indicates that data lie somewhere in between the log-normal distribution and the Weibull one.
The likelihood ratio test has been used to discriminate between log-normal and Weibull distributions, in comparison with the chi-square goodness of the fit~\cite{Cohen65,Dumonceaux73,Dumonceaux73_1,Kundu04,Dey09}.
In those works the likelihood test method is in general favored. Therefore we present data with the likelihood test method rather than least-squares fitting.
Typical data together with fitted distributions via various methods are exhibited in Figs.~\ref{f:strength_chi_like} to \ref{f:weight}: the strength distribution $f(s)$ in Figs.~\ref{f:strength_chi_like}, \ref{f:strength_lognormal}, and \ref{f:strength_Weibull} on the linear scale; the weight distribution $f(w)$ in Figs.~\ref{f:weight_chi_like} and \ref{f:weight} on the logarithmic scale.

\begin{table}[]
\scriptsize
\centering
\renewcommand{\arraystretch}{1.5}
\begin{tabular}{c c c | c c c | c c}
 & & & \multicolumn{3}{c}{Strength} & \multicolumn{2}{|c}{Weight} \\
 & & & {Arrival} & {Departure} & {Total} & {Directed} & Non-directed \\
 \hline
\multirow{4}*{$\ $Morning$\ $} & \multirow{2}*{$\chi^2$} & $\ $Log-normal $\ $
	& $\  2.02 \times 10^{-5}\  $  & $\  2.95 \times 10^{-5}\  $ & $\  7.61 \times 10^{-5}\  $
	& $\  6.39 \times 10^{-8}\  $ & $\ 2.06 \times 10^{-8}\ $ \\
																 &  & Weibull & $3.25 \times 10^{-5}$ & $2.77 \times 10^{-5}$ & $8.02 \times 10^{-5}$
	& $2.90 \times 10^{-7}$ & $9.16 \times 10^{-8}$ \\
 \cline{2-8}
 & \multirow{2}*{$\ln{L}$} & Log-normal & -3324.7 & -3367.9 & -3597.6 & -301608 & -208568 \\
									 & & Weibull & -3354.0 & -3363.5 & -3597.5 & -314671 & -214148 \\
 \hline
\multirow{4}*{Daytime} & \multirow{2}*{$\chi^2$} & Log-normal & $4.98 \times 10^{-5}$ & $4.61 \times 10^{-5}$ & $5.83 \times 10^{-5}$
	& $1.03 \times 10^{-7}$ & $2.29 \times 10^{-8}$ \\
 																&  & Weibull & $5.00 \times 10^{-5}$ & $6.12 \times 10^{-5}$ & $7.83 \times 10^{-5}$
 & $5.96 \times 10^{-7}$ & $1.83 \times 10^{-7}$ \\
 \cline{2-8}
 & \multirow{2}*{$\ln{L}$} & Log-normal & -3208.2 & -3222.3 & -3462.2 & -251962 & -168671 \\
									 & & Weibull & -3218.9 & -3230.2 & -3472.6 & -265075 & -175212 \\
 \hline
\multirow{4}*{Evening} & \multirow{2}*{$\chi^2$} & Log-normal & $2.79 \times 10^{-5}$ & $2.10 \times 10^{-5}$ & $3.75 \times 10^{-5}$
	& $2.08 \times 10^{-8}$ & $6.78 \times 10^{-9}$ \\
																 &  & Weibull & $2.60 \times 10^{-5}$ & $2.98 \times 10^{-5}$ & $4.13 \times 10^{-5}$
	& $1.29 \times 10^{-7}$ & $5.32 \times 10^{-8}$ \\
 \cline{2-8}
 & \multirow{2}*{$\ln{L}$} & Log-normal & -3470.3 & -3428.2 & -3700.3 & -340951 & -222472 \\
									 & & Weibull & -3473.9 & -3456.0 & -3713.1 & -354255 & -228489 \\
 \hline
\multirow{4}*{All day} & \multirow{2}*{$\chi^2$} & Log-normal & $3.54 \times 10^{-5}$ & $4.26 \times 10^{-5}$ & $5.65 \times 10^{-5}$
	& $5.20 \times 10^{-9}$ &  $2.13\times 10^{-9}$ \\
																 &  & Weibull & $3.88 \times 10^{-5}$ & $4.21 \times 10^{-5}$ & $6.02 \times 10^{-5}$
	& $3.18 \times 10^{-8}$ & $1.94 \times 10^{-8}$ \\
 \cline{2-8}
 & \multirow{2}*{$\ln{L}$} & Log-normal & -3745.0 & -3744.5 & -3991.8 & -477253 & -286044 \\
									 & & Weibull & -3755.0 & -3750.8 & -4000.4 & -488448 & -290855
\end{tabular}
\caption{Obtained values of chi square and log of the likelihood function}
\label{t:chi_lnL}
\end{table}

In most cases, log-normal distributions describe the data better than Weibull distributions,
except for the total and departure strength data in the morning, arrival strength data in the evening, and all-day departure strength data.
Whereas the strength data of morning departure fit better to the Weibull distribution,
other data lie somewhere in between the log-normal distribution and the Weibull one.
We therefore choose the log-normal distribution, and compute the growth factor $b$ and the growth frequency $\lambda t$ from Eq.~(\ref{eq:sigma}), which are presented in Table.~\ref{t:a_lt}.
In general larger values of $b$ indicate that the elements (stations or links) have gone through larger changes in each event, while the total number of events occurred is given by $\lambda t$.

\begin{table}[]
\scriptsize
\centering
\renewcommand{\arraystretch}{1.5}
\begin{tabular}{c c | c c c | c c }
\label{growth factor and growth rate}
 & & \multicolumn{3}{c}{Strength} & \multicolumn{2}{c}{Weight} \\
 & & Arrival & Departure & Total & Directed & Non-directed \\
\hline
\multirow{4}*{Morning} & $\mu$ &7.828 & 8.075 & 8.767 & 1.735 & 2.236 \\
& $\sigma$ &$\quad$1.068$\quad$&$\quad$0.941$\quad$&$\quad$0.897$\quad$&$\quad$1.408$\quad$&$\quad$1.509$\quad$\\
& $b$ & 0.157 & 0.116 & 0.096 & 2.136 & 1.769\\
& $\lambda t$ & 53.70 & 73.61 & 95.58 & 1.518 & 2.195\\
\hline
\multirow{4}*{$\ $Afternoon$\ $} & $\mu$ & 7.507 & 7.630 & 8.269 & 1.509 & 1.886\\
& $\sigma$ & 1.062 & 0.977 & 1.010 & 1.128 & 1.427\\
& $\ $ $b$ $\ $ & 0.162 & 0.133 & 0.131 & 1.978 & 1.943\\
& $\lambda t$ & 49.98 & 60.98 & 67.01 & 1.383 & 1.747\\
\hline
\multirow{4}*{Evening} & $\mu$ & 8.326 & 8.122 & 8.964 & 1.895 & 2.365\\
& $\sigma$  & 0.976 & 1.064 & 0.982 & 1.450 & 1.572\\
& $b$ & 0.121 & 0.149 & 0.114 & 2.034 & 1.842\\
& $\lambda t$ & 72.70 & 58.31 & 83.25 & 1.708 & 2.264\\
\hline
\multirow{4}*{All day} & $\mu$ & 9.100 & 9.138 & 9.815 & 2.506 & 3.046\\
& $\sigma$  & 0.971 & 0.934 & 0.949 & 1.586 & 1.694\\
& $b$ & 0.109 & 0.100 & 0.096 & 1.727 & 1.565\\
& $\lambda t$ & 87.74 & 95.78 & 106.93 & 2.498 & 3.234\\
\end{tabular}
\caption{Growth factor and growth frequency calculated from the mean and deviation parameters $(\mu, \sigma)$}
\label{t:a_lt}
\end{table}

%
It is evident that arrival in the morning and departure in the evening are mostly related to stations in downtown or business districts of the city.
Conversely, arrival in the evening and departure in the morning are mainly associated with the residential areas and suburbs of the city.
In Table~\ref{t:a_lt} distributions of morning arrival and evening departure strengths exhibit relatively large values of $b$ but smaller values of $\lambda t$; this indicate that the downtown areas have suffered fewer number of changes, each of which is, however, more drastic.
On the contrary, from morning departure and evening arrival data,
it is concluded that the suburbs and residential areas have experienced gradual changes with
larger frequency.
It is also of interest to note that afternoon data represent the distributions evolved most drastically.
This may be attributed to subway stations close to such facilities as major transportation centers, university towns, and tourist spots, which attract more passengers in the afternoon.
Indeed a few such stations, including particularly the station at the express bus terminal,
are found to have rather large afternoon weights.

In the case of the weight distribution, the log-normal distribution provides better fits
at all times and for both directed and non-directed data.
As expected, the afternoon data show most drastically evolved distributions,
similarly to the strength distribution.
On the other hand, for the strength distribution,
although most cases also follow manifestly log-normal distributions, there are several cases displaying some characteristics of Weibull distributions as well.
Those cases lying in between log-normal and Weibull distributions include strength distributions
of morning departure, evening arrival, morning total, and all-day departure,
among which the former two, morning departure and evening arrival,
constitute the most conspicuous examples.

\section{Discussion}
\label{sec:summary}

We have proposed the master equation approach to the evolution of passengers in a subway system.
With the transition rate constructed from simple geographical consideration, we have obtained the evolution equation for the distribution of subway passengers, which has been found to admit skew distributions including log-normal and Weibull distributions.
Then the Metropolitan Seoul Subway (MSS) system has been considered and the huge trip
data of all passengers in a day have been analyzed by time zones,
which reveals that the passenger trip data in most cases fit well to the log-normal
distributions.
In particular, it has been manifested that while the suburbs have been developed gradually and continually, the downtown areas have undergone rapid development intermittently.
This reflects the trend that the evolution of the city depends mainly upon the development of a few downtown areas whereas the population is distributed to various suburbs and residential areas.
Further, the afternoon data exhibit most drastic evolution, resulting mainly from subway stations
close to such facilities as train stations, bus terminals, university towns, and tourist spots. Construction of such major facilities is thus expected to have a large effect on the passenger trip, and the possibility of swelling a number of passengers drastically should be considered.

Unlike most cases displaying log-normal distributions, a few cases, particularly, the strength distributions of morning departure and evening arrival, follow more or less Weibull distributions.
According to the result of Sec.~\ref{sec:master_eq} that the branching process
gives rise to the Weibull distribution, we may understand emergence of the Weibull distribution as follows:
Presumably, construction of new subway lines will not affect much the passengers using the subway stations in the downtown areas. Accessibility to those areas, high enough, is already secured, and the establishment of additional subway stations may not affect immediately the movement to the downtown. Rather, an increase in passengers in the downtown areas and business districts of the city is largely related to the growth of the city itself and due to the influx from other regions or cities. The growth of the corresponding stations or links may then be regarded as independent of each other. On the other hand, passenger distributions in the suburbs or residential areas are expected to exhibit different behavior. To begin with, construction of new subway stations should stimulate the movement between the catchment areas of the stations. For example, citizens having accessed to the MSS indirectly by means of other transportation means like shuttle buses to a pre-existing major transportation station, can now access to the MSS directly via new subway stations. These redistributions of the population and passengers bear resemblance to branching. To be specific, upon construction of a new subway station, the passengers using the pre-existing station are now divided into two parts: one continuing to use the pre-existing station and the other using the newly constructed station. It is usually more flexible for people to move residences than to move workplaces, which is reflected in the construction of new subway lines and stations. Accordingly, the distributions related to the residential areas, i.e., morning departure or evening arrival, have the origin, at least partially, in the branching process at least partially, and thus the characteristics of the Weibull distribution.

The above discussion allows one to draw a conclusion on the growth of the city. The city first suffers rapid and drastic but infrequent growth as the business districts develop by absorbing populations from other regions.
This is followed by gradual and continuous redistributions of the populations throughout the residential areas.
The master equation approach in this paper makes clear these and helps one to understand the nature of the growth of Seoul City as well as the MSS.

It is also of interest to notice the characteristic of the weight distribution
depending on the time zone.
While the afternoon data, like the strength distribution, show most drastically evolved distributions, the evening data show most gradually evolved distributions.
Such behavior of the evening data reflects the trend that passengers diffuse more in the evening than in the morning, to other places such as entertainment districts or part-time workplaces \cite{metro}.
This tendency is also reflected by the MST structure and the resulting exponent $\gamma$
of the degree distribution. The larger value of the exponent implies that passenger
flows are concentrated on a few nodes (corresponding to downtown business districts).
In contrast passenger flows diffusing to relatively many nodes result in less biased distributions and thus smaller values of the exponent.
This explains the observation that the exponent takes a larger value in the morning ($\gamma = 2.0$)
and a smaller value in the evening ($\gamma = 1.7$).

The time evolution of the distribution of passengers may explicitly probed by analyzing
time series data of the passengers in the subway system, which is left for further study.
Despite the lack of such data sets in this work, we can still conclude that the suburbs have been undergone gradual development, i.e., small developmental changes rather frequently and the downtown areas have passed through rapid development, i.e., large changes once in a while.
This reflects the fact that the evolution of a city is driven mainly by the development of a few downtown areas but the population is distributed in the suburbs and various residential areas
of the city.

\ack
This work was supported in part by the National Research Foundation,
through Grant KRF-2008-327-B00821 and through the Basic Science Research Program (2009-0080791).


\end{document}